\documentclass[useAMS,usenatbib]{mn2e}

\usepackage{graphicx}

\title[Evidence for a lost population of close-in exoplanets]{Evidence for a lost population of close-in exoplanets}
\author[Timothy A. Davis and Peter J. Wheatley]{Timothy A. Davis$^{1,2}$\thanks{E-mail: timothy.davis@astro.ox.ac.uk} and Peter J. Wheatley$^{1}$\thanks{E-mail: p.j.wheatley@warwick.ac.uk}\\
$^{1}$ Department of Physics, University of Warwick, Gibbet Hill Road, Coventry, CV4 7AL\\ $^{2}$ Sub-Department of Astrophysics, University of Oxford, Denys Wilkinson Building, Keble Road, Oxford, OX1 3RH}
\begin{document}

\date{Accepted 2009 March 10.  Received 2009 March 9; in original form 2008 December 22}

\pagerange{\pageref{firstpage}--\pageref{lastpage}} \pubyear{2009}

\maketitle

\label{firstpage}

\begin{abstract}
We investigate the evaporation history of known transiting exoplanets in order to consider the origin of observed correlations between mass, surface gravity and orbital period. We show that the survival of the known planets at their current separations is consistent with a simple model of evaporation, but that many of the same planets would not have survived closer to their host stars. These putative closer-in systems represent a lost population that could account for the observed correlations. We conclude that the relation underlying the correlations noted by \cite{mazeh} and \cite{pjw} is most likely a linear cut-off  in the $M^2/R^3$ vs $a^{-2}$ plane, and we show that the distribution of exoplanets in this plane is in close agreement with the evaporation model. 
\end{abstract}


\section{Introduction}

When \cite{mayor} discovered the first planet around a main sequence star, its small orbital separation came as a surprise to the astronomical community, who had expected to find planets similar to those in our solar system. This motivated work into how and where these `hot Jupiters' form and how they might migrate to their current separations \citep[see for example][]{mig1,mig2}. 

Many close-in planets are now known and intriguingly some planet properties seem to be correlated. \cite{mazeh} suggest that planet mass is correlated with orbital period, while \cite{pjw} find that surface gravity may be correlated with orbital period.  These correlations have been re-plotted by various authors as more planets have been discovered \citep[for example][]{update1,update2}. Both distributions are reproduced in our Figure \ref{correlations}, using the sample of transiting planets from this work (see Section \ref{sampleage}).

It has been suggested that these correlations might be a result of the migration mechanism (e.g. \cite{migr}  and references therein) or that they are a result of evaporation of close-in planets, such as that reported to be ongoing in HD209458b by \cite{vidal} \citep[see also][]{benjaffel,evapagain}.

\cite{lecav} developed a simple model of exoplanet evaporation driven by X-ray/EUV radiation and considered the current evaporation rates of known planets. He found that known planets will not be losing
significant mass at today's irradiation rates. \cite{penzXevap} pointed out that X-ray/EUV irradiation will be much stronger around young stars and showed that, in principle, the mass distribution of close-in exoplanets could be significantly modified by evaporation.

In this paper we use a simple model of evaporation to investigate whether the trends in planet properties identified by \cite{mazeh} and \cite{pjw} can be accounted for by past evaporation. We find good agreement between the distribution of known planets and the cut-off predicted by the evaporation model, and we take this as evidence that a population of close-in exoplanets has been lost to evaporation. 

\section{Method}

\subsection{Selection of exoplanet sample}
\label{sampleage}
We select only transiting planets for our study because we require the planetary radius in order to consider surface gravity and evaporation rates.  We selected planets from The Exoplanets Encyclopedia\footnote{http://www.exoplanet.eu} on 15 June 2008. Key parameters were selected from the encyclopedia and, where necessary, original discovery papers. These parameters allow the calculation of the energy incident upon the planet per unit time, and the total potential energy of the planet. Stellar quantities are shown in Table \ref{table2}, and planetary properties in Table \ref{table1} for the sample of planets used in this work.

We used as many of the known transiting systems as possible, although some were removed due to the absence of data required in our analysis. The CoRoT systems and Lupus-TR-3b were removed due to lack of distance information available at the time of our analysis. OGLE-TR-10b, 111b, 182b, 211b, and SWEEPS-04b and 11b were removed due to the host stars being of unknown spectral type. GJ 436b was not included because it has the only M type host, and will have different high energy emisison characteristics to the FGK hosts prevalent in the rest of the sample. HD149026b was not included because its host star is not on the main sequence. 

\begin{figure}
 \begin{center}
 \includegraphics[scale=0.55]{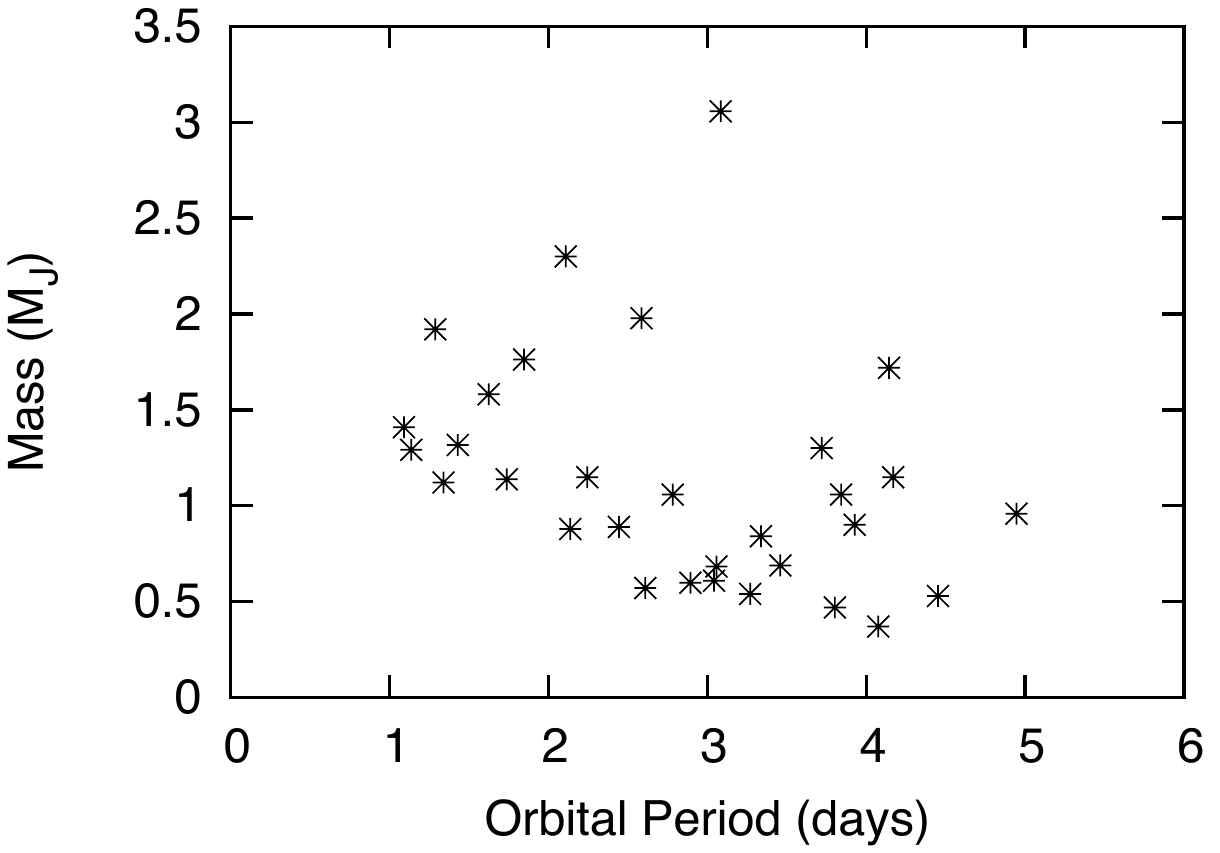}
 \hfill
 \includegraphics[scale=0.55]{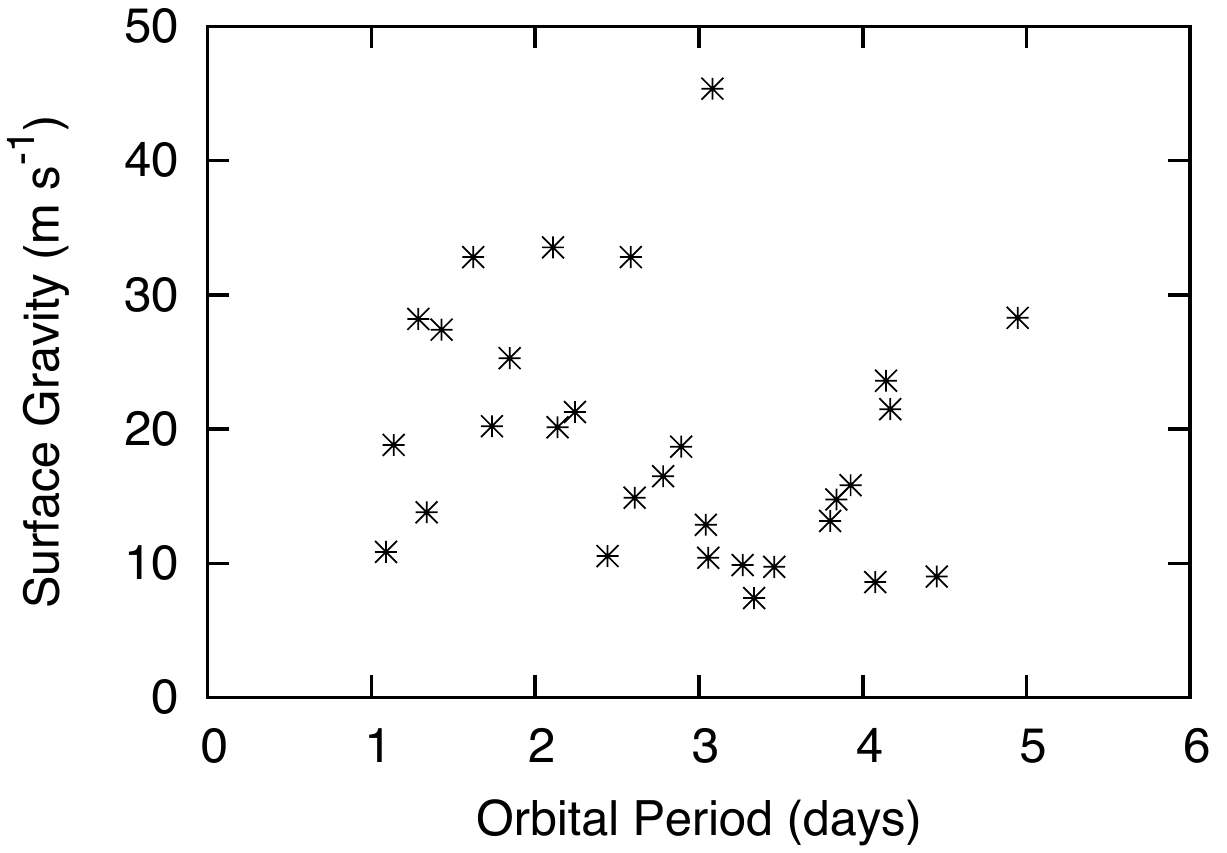}
 \caption{Plots of the property correlations of {\protect \cite{mazeh}} and {\protect \cite{pjw}}, updated to include the entire sample of planets as listed in Table {\protect \ref{table1}}. Planets exist beyond the area of the plot, specifically HAT-P-2b, but always towards the top right.}
 \label{correlations}
 \end{center}
\end{figure}

In Figure \ref{correlations} we plot mass against orbital period for our sample after \cite{mazeh}, and surface gravity vs period after \cite{pjw}. Both plots show greater scatter than the original suggested correlations, especially to the upper right, but the mass-period distribution still shows a sharp cut-off to the lower left. The lower left of the surface gravity-period distribution does not show a sharp cutoff, but this area remains relatively sparsely populated, despite strong selection effects in transit surveys for planets in this region. The upper right of both of these diagrams also remain
sparsely populated. This is unlikely to be related to evaporation and instead may be a feature of planetary migration. Selection effects in transit surveys may also be a factor, since they select against long orbital periods and high surface gravities.

\begin{table*}
\centering
\begin{minipage}{135mm}
\caption{Stellar parameters used in this work. Stellar mass, radius and spectral type are taken from The Exoplanet Encyclopedia$^a$ unless otherwise noted, and bolometric magnitudes from \protect \cite{astroquant}. $\rho_*$ is calculated using the standard formula from the data listed. $L_x^{sat}$ is calculated by converting the bolometric magnitude into a bolometric luminosity, and using the relations from \protect \cite{pizzolato} as described in Section \protect \ref{xrayest}. $\tau_{sat}$ is estimated as described in Section \protect \ref{xrayest}.}
\begin{tabular}{c c c c c c c c c c}
\hline
Star & Mass & Radius & $\rho_*$ & Spec.Type & Bol.Mag. & $\log{L_x^{sat}}$& $\tau_{sat}$ & ref\\
 & \small ($M_{\odot}$) & \small ($R_{\odot}$) & g cm$^{-3}$ & &\small (mag)& \small (ergs s$^{-1}$)& \small (Gyr) &\\
\hline
HAT-P-1  &1.12&1.15&1.04& GOV  &4.2&29.9&0.2& \small a\\ 
HAT-P-2  &1.298&1.41&0.65& F8  &3.8&30.06&0.1& \small a\\ 
HAT-P-3  &0.936&0.82&2.36& K  &5.6&30.14&0.35& \small a\\ 
HAT-P-4  &1.26&1.59&0.44& F7V  &3.5&30.18&0.1& \small a\\ 
HAT-P-5  &1.16&1.17&1.03& G1V  &4.5&29.78&0.1& \small a\\ 
HAT-P-6  &1.29&1.46&0.59& F1V  &3.2&30.3&0.1& \small a\\ 
HD 17156  &1.2&1.47&0.53& GO  &4.2&29.9&0.2& \small a\\ 
HD 189733  &0.8&0.75&2.64& K1-K2  &5.6&30.04&0.35& \small a\\ 
HD 209458  &1.01&1.12&1.01& GOV  &4.2&30.6&0.2& \small a\\ 
OGLE-TR-113  &0.78&0.77&2.41& K  &5.6&30.14&0.35& \small a\\ 
OGLE-TR-132  &1.26&1.34&0.74& F5  &3.5&30.18&0.1& \small a\\ 
OGLE-TR-56  &1.17&1.32&0.72& G0  &4.2&29.9&0.2& \small a\\ 
TrES-1   &0.87&0.82&2.23& KOV  &5.6&30.04&0.35& \small a\\ 
TrES-2   &0.98&1.00&1.38& GOV  &4.2&30.7&0.2& \small a\\ 
TrES-3   &0.924&0.81&2.43& G6V  &5&30.38&0.2& \small a,b\\
TrES-4   &1.384&1.81&0.33& F8V  &3.8&30.06&0.1& \small a\\ 
WASP-1  &1.24&1.38&0.66& F7V  &3.65&30.12&0.1& \small a,c\\
WASP-2  &0.84&0.83&2.04& K1V  &5.7&30.00&0.35& \small a,c\\
WASP-3  &1.24&1.31&0.78& F7V  &3.65&30.12&0.1& \small a,c\\
WASP-4  &0.9&1.15&0.84& G7V  &5&30.38&0.2& \small a,c\\
WASP-5  &0.972&1.03&1.27& G4V  &4.8&30.46&0.2& \small a,c\\
WASP-6  &0.85&0.9&1.65& G8  &5.1&30.24&0.2& \small a,c\\
WASP-7  &1.28&1.24&0.96& F5V  &3.5&30.18&0.1& \small a,c\\
WASP-8  &0.87&0.91&1.63& G6  &5&30.28&0.2& \small a,c\\
WASP-9  &1.1&1.05&1.34& G0  &4.2&30.6&0.2& \small a,c\\
WASP-10  &0.71&0.78&2.09& K5  &6.7&29.7&0.35& \small a,c\\
WASP-11  &0.74&0.74&2.58& K3  &6.1&29.94&0.35& \small a,c\\
WASP-12  &1.1&1.05&1.34& G0  &4.2&30.6&0.2& \small a,d\\
WASP-13  &1.3&1.3&0.84& F9  &4&29.98&0.1& \small a,c\\
WASP-14  &1.319&1.3&0.85& F5V  &3.5&30.18&0.1& \small a,c\\
WASP-15  &1.4&1.4&0.72& F5  &3.5&30.18&0.1& \small a,c\\
XO-1  &1.00&0.93&1.77& G1V  &4.3&30.56&0.2& \small a\\ 
XO-2  &0.98&0.96&1.54& KOV  &5.6&30.14&0.35& \small a\\
XO-3  &1.213&1.38&0.66& F5V  &3.5&30.18&0.1& \small a\\
XO-4  &1.32&1.55&0.5& F5V  &3.5&30.18&0.1& \small a\\
XO-5  &1.00&1.11&1.03& G8V  &5.1&30.24&0.2& \small a\\
\hline
\label{table2}
\end{tabular}
\\\emph{References:}
$^a$ http://www.exoplanet.eu - retrieved 15/JUN/08. $^b$ \cite{odon}. $^c$ \cite{cameron} $^d$ \cite{hebb}.
\end{minipage}
\end{table*}

\begin{table*}
\centering
\begin{minipage}{140mm}
\caption{Planetary parameters used in this work, includes calculated figures for $\langle$a$\rangle$, a$_{roche}$, PE and $\langle\frac{dE}{dt}\rangle$. Figures for a, R$_{p}$ and M$_{p}$ are taken from The Exoplanet Encyclopedia$^a$ unless otherwise specified. Densities may be compared with that of Jupiter, 1.33 g cm$^{-3}$ \protect \citep[][]{astroquant}.}
\begin{tabular}{cccccccccc}
\hline
Planet&$<$a$>$&a$_{roche}$&M$_p$&R$_p$&$\rho_p$&$\log(-PE)$&$\log\langle\frac{dE}{dt}\rangle_{sat}$&$\log\langle\frac{dE}{dt}\rangle_{roche}$&Ref.\\ 
\tiny &(au) & (au) & (M$_j$) & (R$_j$) & (g cm$^{-3}$) & (ergs) & (ergs $\tau_{sat}^{-1}$) & (ergs $\tau_{sat}^{-1}$ ) & \\
\hline
HAT-P-1b&0.055&0.018&0.53&1.203&0.38&42.77&42.33&43.29& \small a\\
HAT-P-2b&0.077&0.006&8.64&0.952&12.42&45.29&41.1&43.21&\small a\\
HAT-P-3b&0.039&0.012&0.599&0.89&1.05&43&42.44&43.45&\small a\\
HAT-P-4b&0.045&0.018&0.68&1.27&0.41&42.96&42.75&43.51&\small a\\
HAT-P-5b&0.041&0.015&1.06&1.26&0.66&43.35&42.62&43.47&\small a\\
HAT-P-6b&0.052&0.017&1.057&1.33&0.56&43.32&42.67&43.65&\small a\\
HD 17156b&0.184&0.01&3.08&1.15&2.51&44.31&40.65&43.01&\small a\\
HD 189733b&0.031&0.012&1.15&1.154&0.93&43.46&42.26&43.09&\small a\\
HD 209458b&0.045&0.018&0.69&1.32&0.37&42.96&42.79&43.6&\small a\\
OGLE-TR-113b&0.023&0.011&1.32&1.09&1.26&43.6&42.48&43.13&\small a\\
OGLE-TR-132b&0.031&0.014&1.14&1.18&0.86&43.44&42.21&42.86&\small a\\
OGLE-TR-56b&0.023&0.015&1.29&1.3&0.73&43.51&42.58&42.94&\small a\\
TrES-1b&0.039&0.014&0.61&1.081&0.6&42.93&42.6&43.47&\small a\\
TrES-2b&0.037&0.011&1.98&1.22&1.35&43.91&42.1&43.12&\small a\\
TrES-3b&0.023&0.012&1.92&1.295&1.1&43.85&42.25&42.8&\small a,b\\
TrES-4b&0.049&0.023&0.84&1.674&0.22&43.02&42.69&43.33&\small a,c\\
WASP-1b &0.038&0.019&0.89&1.443&0.37&43.14&42.93&43.53&\small a,c\\
WASP-2b &0.031&0.012&0.88&1.038&0.98&43.27&42.95&43.75&\small a,c\\
WASP-3b &0.032&0.014&1.76&1.31&0.97&43.77&42.21&42.93&\small a,c\\
WASP-4b &0.023&0.016&1.1215&1.416&0.49&43.35&42.32&42.65&\small a,c\\
WASP-5b &0.027&0.011&1.58&1.09&1.51&43.76&42.04&42.81&\small a,c\\
WASP-6b &0.045&0.006&1.3&0.5&12.8&43.93&40.8&42.61&\small a,c\\
WASP-7b &0.062&0.012&0.96&0.915&1.55&43.4&42.18&43.61&\small a,c\\
WASP-8b &0.082&0.011&2.35&1.21&1.64&44.06&41.07&42.81&\small a,c\\
WASP-9b &0.033&0.011&2.69&1.22&1.83&44.17&42.19&43.14&\small a,c\\
WASP-10b &0.037&0.009&3.06&1.29&1.76&44.26&41.77&42.96&\small a,c\\
WASP-11b &0.043&0.013&0.58&0.98&0.76&42.93&42.24&43.3&\small a,c\\
WASP-12b &0.0229&0.021&1.41&1.79&0.3&43.44&42.84&42.92&\small a,d\\
WASP-13b &0.055&0.017&0.49&1.06&0.51&42.75&42.01&43.04&\small a,c\\
WASP-14b &0.036&0.008&7.725&1.259&4.78&45.07&42.13&43.4&\small a,c\\
WASP-15b &0.048&0.02&0.56&1.33&0.42&42.77&42.52&43.3&\small a,c\\
XO-1b&0.049&0.015&0.9&1.184&0.67&43.23&42.59&43.64&\small a\\
XO-2b&0.037&0.014&0.57&0.973&0.77&42.92&42.57&43.42&\small a\\
XO-3b&0.047&0.007&11.79&1.217&8.08&45.46&41.87&43.52&\small a\\
XO-4b&0.056&0.015&1.72&1.34&0.88&43.74&41.8&42.97&\small a\\
XO-5b&0.051&0.013&1.15&1.15&0.93&43.46&41.41&42.59&\small a\\
\hline
\label{table1}
\end{tabular}
\emph{References:}
$^a$ http://www.exoplanet.eu - retrieved 15/JUN/08. $^b$ \cite{odon}.  $^c$ \cite{cameron} $^d$ \cite{hebb}.
\end{minipage}
\end{table*}

\subsection{Evaporation model}
\label{evapmod}
It has been shown by \cite{lammer} that Jeans escape of material from hot Jupiters cannot drive significant mass loss. Instead they suggest that energy-limited escape of particles due to absorption of EUV/X-ray photons is the most-likely driver of mass loss in these systems. 

In order to estimate the evaporation rates of exoplanets we adopt the model of \cite{lecav}, who used
the ratio of planetary potential energy and the time-averaged power incident on the planet in order to predict mass loss rates, given by

\begin{equation}
\dot{m} = \frac{4L_{x}R_{p}^{3}}{3G M_{p} \langle a\rangle^{2}}
\label{mloss}
\end{equation}
where $\dot{m}$ is mass loss rate, $L_{x}$ is the star's X-ray/EUV luminosity, $R_{p}$ is the planetary radius, $M_{p}$ is the planetary mass, G is the gravitational constant, and $\langle$a$\rangle$ is the time-averaged orbital distance.

This simple model assumes energy-limited evaporation, which may overestimate real mass loss rates. It also assumes that a test particle must be moved to infinity to escape the gravitational potential of a planet \citep[Sect.\,4.3 of][]{lecav}, which \cite{rochelobeeffect} suggest may underestimate mass loss by up to 40\% when a planet is close to filling its Roche lobe. Further, we neglect any evolution of the planet radius, assuming the current value for each planet. Since unperturbed gaseous planets contract slowly with time, this will tend to underestimate mass loss rates.

\subsection{Estimation of X-ray/EUV irradiation}
\label{xrayest}

X-ray/EUV emission from stars is believed to be driven by a dynamo action that decreases with time as the star spins down due to magnetic braking \citep{Skumanic}. X-ray/EUV emission is seen to increase with stellar rotation rate, up to a maximum of L$_x$/L$_{bol}$ $\simeq$ 10$^{-3}$, where it saturates  \citep{vil,vilwil}. 
We expect therefore all known exoplanets to have suffered a period of high and approximately constant X-ray/EUV irradiation in the past, when their host stars were young and rapidly rotating.  This saturated phase is likely to dominate their irradiation history. 

The duration of the saturated phase of X-ray emission depends on spectral type, but is not well known. For F type stars a saturation time of 0.1 Gyr is used, taken from \cite{fstarsat} who used optical and X-ray telescope data of five star-forming regions of varying age ($\rho$ Ophiuchi, the Orion Nebula Cluster, NGC 2264, Chamaeleon I, and $\eta$ Chamaeleontis) and two young clusters (the Pleiades 
and NGC 2516). This is their upper limit for the saturation period of stars with the mass of F stars. 

For G type stars we estimate a saturation time of 0.2 Gyr using the results of \cite{gstarsat}, who used the ASCA and ROSAT X-ray satellites to probe the coronae of a sample of nine solar-like G stars, with ages between 70 Myr and 9 Gyr.

The saturation timescale for K type stars is estimated from the observation \citep[e.g., ][]{xraysatwheatley} that K stars are saturated in the Pleiades \citep[age of 0.1 Gyr,][]{pleiades} but not in the Hyades \citep[age of $\sim$0.6 Gyr,][]{hyades}. We choose a value of 0.35 Gyr, but adjusting this value to the extremes does not affect our conclusions (see Section \ref{effect}).

For each of the exoplanets in our sample,  we have calculated the total incident energy over the planet surface for the complete saturation interval of its host star. The stellar parameters required are listed in Table \ref{table2} and planetary properties in Table \ref{table1}. The saturation duration is based only on the spectral type of the star. The X-ray saturation level used can be found by using the saturation level for stars of a given mass \citep[in Table 3 of][]{pizzolato}. Four stars in this work are slightly more massive than those studied by \cite{pizzolato}, and a saturated value of $L_{x}^{sat}/L_{bol} = 10^{-3.9}$ is used, which is the average of the highest mass group in Pizzolato's work. 

A difference between our analysis and the population study of \cite{penzXevap} is that Penz et al. included a range of X-ray behaviour for stars of each type, whereas we assume average properties, since we do not know the individual histories of the objects in our study. Our estimates of X-ray irradiation would be biassed if the X-ray emission of stars were correlated with the presence of planets. Indeed, \cite{excessXRAY} find evidence for excess X-ray emission from stars with known planets. We choose not to modify our estimates by this factor because we are interested in the duration and strength of the saturated phase of X-ray emission, whereas the measurements of Kasyap et al.\ apply to the weaker X-ray emission of much older stars. Nevertheless, if the results of \cite{excessXRAY}  were found to apply to stars of all ages, then we may have underestimated the X-ray irradiation of our sample by around half an order of magnitiude.

\section{Results}
\subsection{Past evaporation of known systems}

\begin{figure}
 \begin{center}
 \includegraphics[scale=0.55]{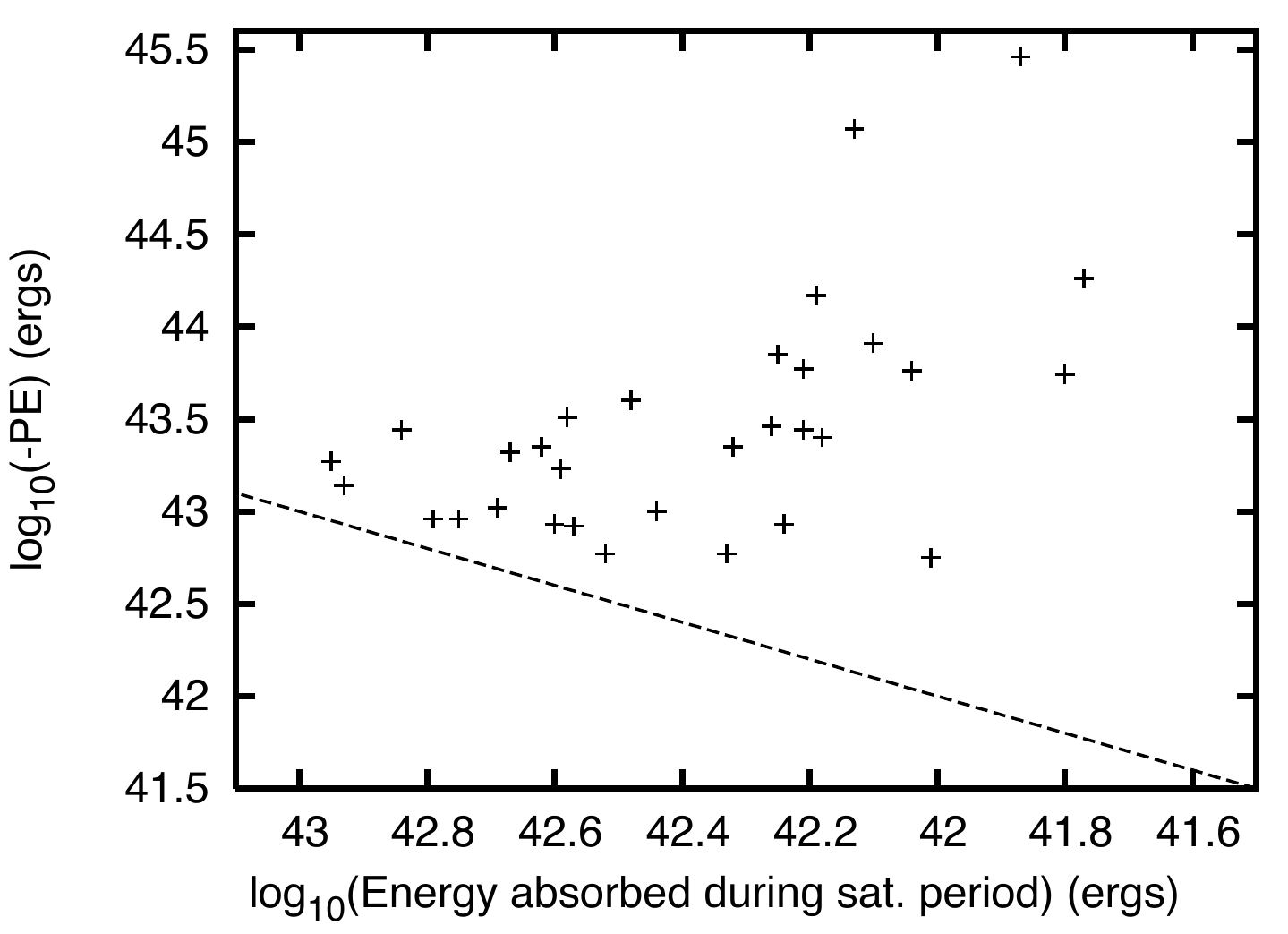}
 \caption{A diagram of potential energy against the X-ray/EUV energy incident on the planet in one saturation period for the known transiting systems in their current configurations. The dashed line is an isocron representing one saturation period.}
 \label{sat}
 \end{center}
\end{figure}

Figure \ref{sat} compares the potential energy of planets in our sample with the energy incident during the saturated X-ray emission phase of their parent star. As can be clearly seen, all of the systems considered in this work are able to survive the energy absorbed during the saturation period of their parent star. Of course, each of these systems has already weathered one saturation period, so Fig. \ref{sat} really tests whether these systems could survive a second saturation period. Since they all could survive another saturation period, it seems likely that they were only minimally affected by the first period. However, systems close to the dashed line may have lost around half of their mass during the saturation period they have aleady weathered.

\subsection{Evaporation at the Roche limit}

\begin{figure}
 \begin{center}
 \includegraphics[scale=0.6]{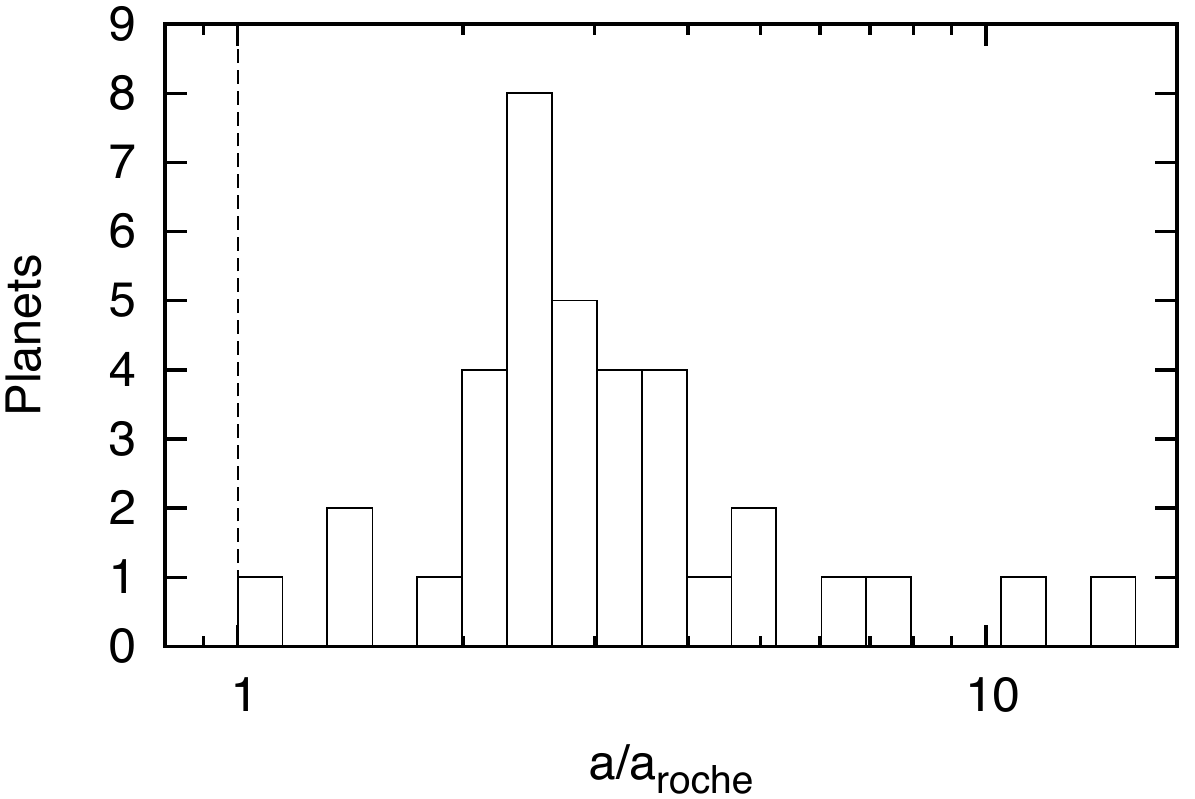}
 \caption{A histogram showing the orbital separation of the sample of exoplanets from Table \ref{table1}, scaled by their Roche limit. The majority of known systems can be moved approximately 2-3 times closer to their parent star.}
 \label{aoveraroche}
 \end{center}
\end{figure}

We know that all the planets in our sample have survived the saturation phase of their star. However most of these systems could have existed closer in, since most are well outside their Roche limit. In fact, there is a clear deficit of systems close to their Roche limits, as shown in Fig. \ref{aoveraroche}, despite selection effects biasing surveys towards detection of close-in systems. The Roche limit sets the minimum orbital separation for a planet, within which tidal forces become too strong and pull the planet apart. 

The Roche limit for a fluid body (gas giant planets are assumed to be best represented by a fluid model rather than a rigid body), as given by \cite{rochelim}, is
\begin{equation}
a_{roche} = 2.44a\log{\left(\frac{\rho_{p}}{\rho_{*}}\right)}
\label{rochelim}
\end{equation} 
where $a_{roche}$ is the Roche limit, a is the semi major axis, and $\rho_{p}$ and $\rho_{*}$ are the density of the planet and star respectively. 

\begin{figure}
 \begin{center}
 \includegraphics[scale=0.55]{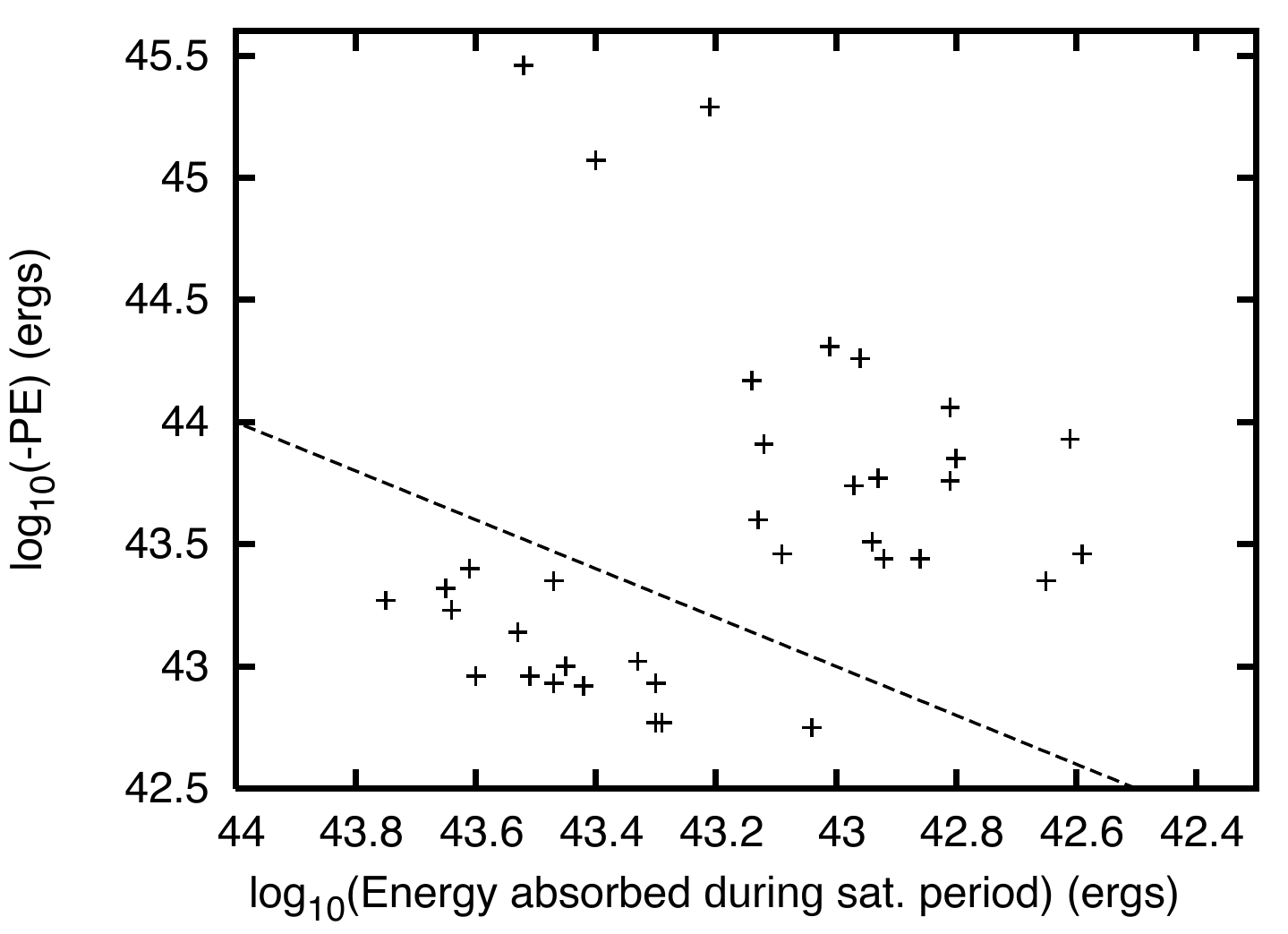}
 \caption{A diagram of potential energy against the X-ray/EUV energy incident on the planet in one saturation period for the known transiting systems, if they existed at their Roche limit. The dashed line is an isocron representing one saturation period.}
 \label{roche}
 \end{center}
\end{figure}

Re-plotting Figure \ref{sat}, but moving the planets in to their Roche limit results in Figure \ref{roche}. In this case $\sim$50\% of known systems would have absorbed enough energy to move their entire mass out to infinity. Therefore closer-orbiting analogs of the known planets would have suffered catastrophic evaporation. The loss of such systems may account for the cut-off seen in Figure \ref{correlations}, since they would be situated to the left of their current positions, thereby filling the sparsely populated areas of these diagrams. Planets that would have survived at their Roche limit tend to have either high densities and surface gravities, or already exist close to their Roche limits.

\subsection{Effect on planet property distributions}
\label{effect}

\cite{mazeh} and \cite{pjw} have suggested that the apparent correlations found in mass-period and surface gravity-period plots for the known planetary systems may be due to evaporation processes. We have shown that if planets like the known sample existed closer to their stars they would not have survived the saturated phase of stellar X-ray/EUV emission, and hence would appear as a void in these plots. 

\begin{figure}
 \begin{center}
 \includegraphics[scale=0.58]{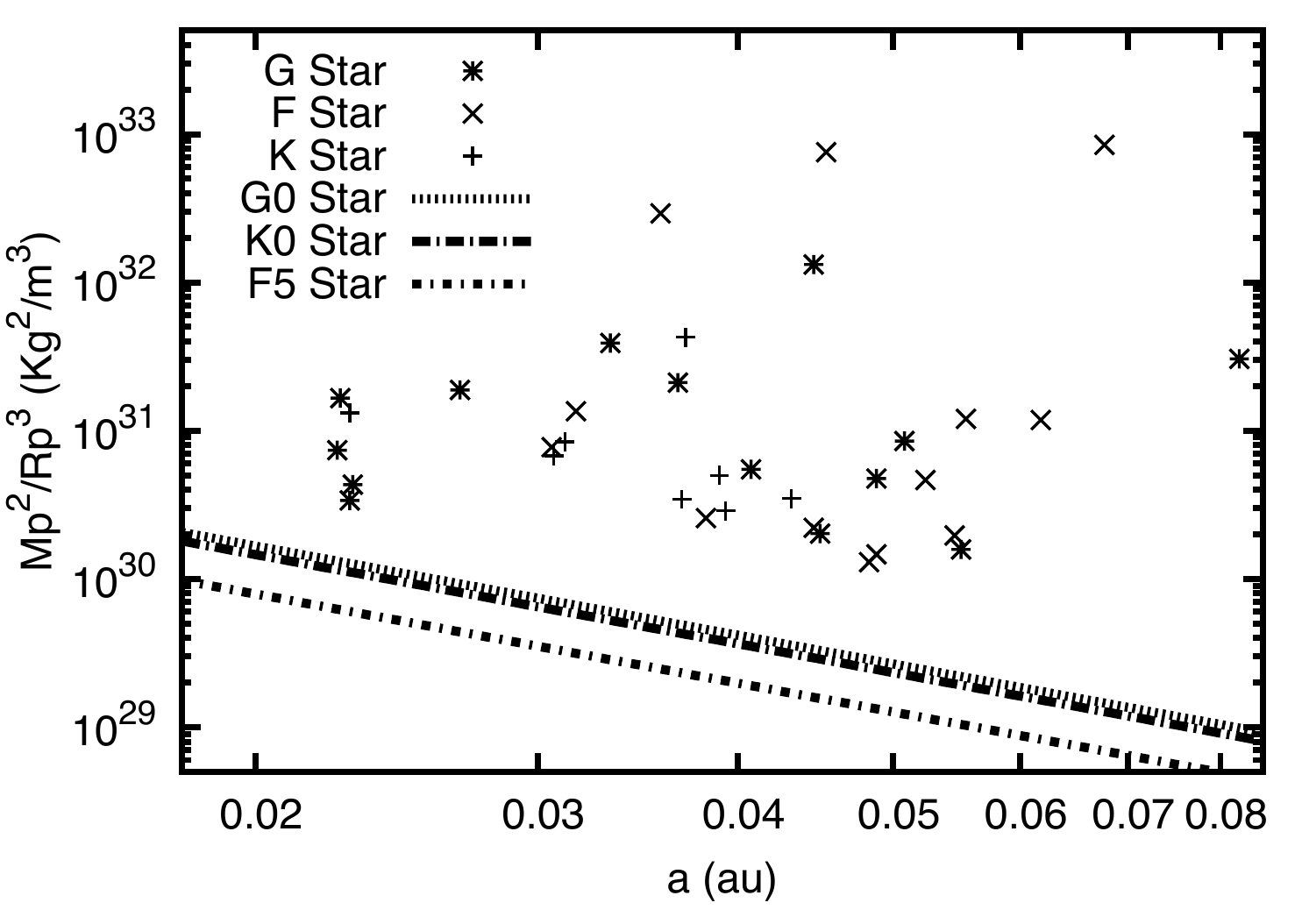}
 \caption{A diagram plotting mass squared over radius cubed versus the inverse of the semi major axis of the planetary orbit squared. Planets that exist around G type stars are represented with star symbols, K type stars are represented by straight crosses, and F type stars by diagonal crosses. The lines represent destruction limits, below which a system would have completely evaporated in the saturation time of its star. The dotted line represents this cutoff for G0 stars, the line with large dashes represents K0 stars, and the line with small dashes represents F5 stars.}
 \label{printcorr}
 \end{center}
\end{figure}

If X-ray/EUV irradiation is important, then the underlying cut-off must be linear in the $M_p^2/R_p^3$ versus $a^{-2}$ plane, as the ratio of potential energy to irradiating energy per unit time defines the planet's characteristic lifetime.  A plot of this form is shown in Figure \ref{printcorr}. A sparsely populated region is seen in the lower left, similar to those in Figure \ref{correlations}. For a given class of star it is possible to plot a power law on this diagram, representing a destruction limit, below which planets would either have evaporated completely or left planetary cores that are not currently detectable in transit surveys. The functional form of this destruction limit is

\begin{equation}
\frac{M_p^2}{R_p^3} = \left(\frac{4}{3}\frac{L_{x}^{sat}\tau_{sat}}{G}\right)a^{-2}
\label{evapline}
\end{equation} 
where M$_p$ is the mass of the planet, R$_p$ is the planetary radius, $a$ is the semi major axis, L$_{x}^{sat}$ is the saturated X-ray/EUV luminosity of the star, $\tau_{sat}$ is the saturation timescale of the spectral type and G is the gravitational constant.

As can be seen in Figure \ref{printcorr}, the plotted lines corresponding to FGK stars track the edge of the missing region very well without any fitting. All systems exist above the destruction line for their spectral type. 

Uncertainties and assumptions in our simple evaporation model (as discussed in Sections \ref{evapmod} and \ref{xrayest}) could move the destruction limits plotted in Fig. \ref{printcorr}., but they are unlikely to change the functional form, which matches the observed distribution well.

The destruction limits can be raised by factors of 8, 3 and 8.5 for F, G and K stars respectively before known planets (in their current state) fall into the danger zone. This suggests that our assumptions have tended to underestimate the effects of evaporation. This is consistent with our expectations, since neglecting Roche geometry, planet radius evolution (Sect. \ref{evapmod}), X-ray irradiation since the saturated phase, and the observed excess X-ray emission of planet hosting stars (Sect. \ref{xrayest}) are all likely to result in underestimates of evaporation rates.

We note that uncertainties in saturation timescales alone cannot move the destruction limits up to the known planets. This would require timescales of 0.8, 0.6 and 3 Gyr for F, G and K stars respectively, which are unreasonably high values compared with existing X-ray observations (Sect. \ref{xrayest}).


\section{Conclusions}
We have shown that a simple model of exoplanet evaporation during the saturated phase of stellar activity is consistent with the observed population of transiting planets. The model also suggests that these same planets would not, in general, have survived at their Roche limit. We conclude that there is likely to have been a population of close-in planets that have been lost to evaporation. If this is the case there is no need for migration models to account for the observed lack of planets close to their Roche limits.

This lost population can account for the correlations between mass-period and surface gravity-period suggested by \cite{mazeh} and \cite{pjw}.  Further to this we suggest that a cut-off in the $M_p^2/R_p^3$ versus $a^{-2}$ plane is the underlying relation, and we show that evaporation in the saturated phase defines destruction limits that are a good match to the position of the lower edge of the observed distribution. 


\section{Acknowledgements}
T. Davis wishes to thank S. D. Smith for his invaluable help and support in the preparation of this paper. The astronomy and astrophysics group at the University of Warwick is supported by a STFC rolling grant. T. Davis is supported by an STFC Postgraduate Studentship. 

\bsp

\bibliographystyle{mn2e}
\bibliography{exoref}

\begin{thebibliography}{}

\bibitem[\protect\citeauthoryear{{Aggarwal} \& {Oberbeck}}{{Aggarwal} \&
  {Oberbeck}}{1974}]{rochelim}
{Aggarwal} H.~R.,  {Oberbeck} V.~R.,  1974, The Astrophysical Journal, 191, 577

\bibitem[\protect\citeauthoryear{Allen}{Allen}{2000}]{astroquant}
Allen C.~W.,  2000, Astrophysical Quantities.
Springer; 4th ed.

\bibitem[\protect\citeauthoryear{{Baraffe}, {Selsis}, {Chabrier}, {Barman},
  {Allard}, {Hauschildt} \& {Lammer}}{{Baraffe} et~al.}{2004}]{barevap}
{Baraffe} I.,  {Selsis} F.,  {Chabrier} G.,  {Barman} T.~S.,  {Allard} F.,
  {Hauschildt} P.~H.,    {Lammer} H.,  2004, Astronomy and Astrophysics, 419,
  L13

\bibitem[\protect\citeauthoryear{{Basri}, {Marcy} \& {Graham}}{{Basri}
  et~al.}{1996}]{pleiades}
{Basri} G.,  {Marcy} G.~W.,    {Graham} J.~R.,  1996, The Astrophysical
  Journal, 458, 600

\bibitem[\protect\citeauthoryear{{Ben-Jaffel}}{{Ben-Jaffel}}{2007}]{benjaffel}
{Ben-Jaffel} L.,  2007, The Astrophysical Journal, 671, L61

\bibitem[\protect\citeauthoryear{{Boss}}{{Boss}}{1996}]{mig2}
{Boss} A.~P.,  1996, in Lunar and Planetary Institute Conference Abstracts
  Vol.~27 of Lunar and Planetary Inst. Technical Report, {Forming a
  Jupiter-like Companion for 51 Pegasi}.
pp 139--+

\bibitem[\protect\citeauthoryear{{Cameron et al.}}{{Cameron et
  al.}}{2008}]{cameron}
{Cameron et al.} 2008 Vol.~253 of Proceedings IAU Symposium, {The WASP transit
  surveys}

\bibitem[\protect\citeauthoryear{{Erkaev}, {Kulikov}, {Lammer}, {Selsis},
  {Langmayr}, {Jaritz} \& {Biernat}}{{Erkaev} et~al.}{2007}]{rochelobeeffect}
{Erkaev} N.~V.,  {Kulikov} Y.~N.,  {Lammer} H.,  {Selsis} F.,  {Langmayr} D.,
  {Jaritz} G.~F.,    {Biernat} H.~K.,  2007, Astronomy and Astrophysics, 472,
  329

\bibitem[\protect\citeauthoryear{Flaccomio, Micela \& Sciortino}{Flaccomio
  et~al.}{2003}]{fstarsat}
Flaccomio E.,  Micela G.,    Sciortino S.,  2003, Astronomy and Astrophysics,
  402, 277

\bibitem[\protect\citeauthoryear{G\"{u}del, Guinan \& Skinner}{G\"{u}del
  et~al.}{1997}]{gstarsat}
G\"{u}del M.,  Guinan E.~F.,    Skinner S.~L.,  1997, The Astrophysical
  Journal, 483, 947

\bibitem[\protect\citeauthoryear{{Hansen} \& {Barman}}{{Hansen} \&
  {Barman}}{2007}]{update1}
{Hansen} B.~M.~S.,  {Barman} T.,  2007, The Astrophysical Journal, 671, 861

\bibitem[\protect\citeauthoryear{{Hebb et al.}}{{Hebb et al.}}{2008}]{hebb}
{Hebb et al.} 2008, (ApJ, In press. arXiv:0812.3240)

\bibitem[\protect\citeauthoryear{{Kashyap}, {Drake} \& {Saar}}{{Kashyap}
  et~al.}{2008}]{excessXRAY}
{Kashyap} V.~L.,  {Drake} J.~J.,    {Saar} S.~H.,  2008, The Astrophysical
  Journal, 687, 1339

\bibitem[\protect\citeauthoryear{{Lammer}, {Selsis}, {Ribas}, {Guinan}, {Bauer}
  \& {Weiss}}{{Lammer} et~al.}{2003}]{lammer}
{Lammer} H.,  {Selsis} F.,  {Ribas} I.,  {Guinan} E.~F.,  {Bauer} S.~J.,
  {Weiss} W.~W.,  2003, The Astrophysical Journal, 598, L121

\bibitem[\protect\citeauthoryear{{Lastennet}, {Valls-Gabaud}, {Lejeune} \&
  {Oblak}}{{Lastennet} et~al.}{2000}]{hyades}
{Lastennet} E.,  {Valls-Gabaud} D.,  {Lejeune} T.,    {Oblak} E.,  2000, in
  {Pallavicini} R.,  {Micela} G.,   {Sciortino} S.,  eds, Stellar Clusters and
  Associations: Convection, Rotation, and Dynamos. Vol.~198 of Astronomical
  Society of the Pacific Conference Series, {Influence of Hipparcos on Hyades
  age estimates from three binary systems}.
p.~133

\bibitem[\protect\citeauthoryear{{Lecavelier Des Etangs}}{{Lecavelier Des
  Etangs}}{2007}]{lecav}
{Lecavelier Des Etangs} A.,  2007, Astronomy and Astrophysics, 461, 1185

\bibitem[\protect\citeauthoryear{{Lin}, {Bodenheimer} \& {Richardson}}{{Lin}
  et~al.}{1996}]{mig1}
{Lin} D.~N.~C.,  {Bodenheimer} P.,    {Richardson} D.~C.,  1996, Nature, 380,
  606

\bibitem[\protect\citeauthoryear{{Mayor} \& {Queloz}}{{Mayor} \&
  {Queloz}}{1995}]{mayor}
{Mayor} M.,  {Queloz} D.,  1995, Nature, 378, 355

\bibitem[\protect\citeauthoryear{{Mazeh}, {Zucker} \& {Pont}}{{Mazeh}
  et~al.}{2005}]{mazeh}
{Mazeh} T.,  {Zucker} S.,    {Pont} F.,  2005, Monthly Notices of the Royal
  Astronomical Society, 356, 955

\bibitem[\protect\citeauthoryear{{O'Donovan}, {Charbonneau}, {Bakos},
  {Mandushev} \& {Dunham}}{{O'Donovan} et~al.}{2007}]{odon}
{O'Donovan} F.~T.,  {Charbonneau} D.,  {Bakos} G.~{\'A}.,  {Mandushev} G.,
  {Dunham} 2007, The Astrophysical Journal Letters, 663, L37

\bibitem[\protect\citeauthoryear{{Penz}, {Micela} \& {Lammer}}{{Penz}
  et~al.}{2008}]{penzXevap}
{Penz} T.,  {Micela} G.,    {Lammer} H.,  2008, Astronomy and Astrophysics,
  477, 309

\bibitem[\protect\citeauthoryear{{Pizzolato}, {Maggio}, {Micela}, {Sciortino}
  \& {Ventura}}{{Pizzolato} et~al.}{2003}]{pizzolato}
{Pizzolato} N.,  {Maggio} A.,  {Micela} G.,  {Sciortino} S.,    {Ventura} P.,
  2003, Astronomy and Astrophysics, 397, 147

\bibitem[\protect\citeauthoryear{{Pollacco}, {Skillen}, {Collier Cameron},
  {Loeillet} \& {Stempels}}{{Pollacco} et~al.}{2008}]{update2}
{Pollacco} D.,  {Skillen} I.,  {Collier Cameron} A.,  {Loeillet} B.,
  {Stempels} H.~C.,  2008, Monthly Notices of the Royal Astronomical Society,
  385, 1576

\bibitem[\protect\citeauthoryear{{Skumanich}}{{Skumanich}}{1972}]{Skumanic}
{Skumanich} A.,  1972, The Astrophysical Journal, 171, 565

\bibitem[\protect\citeauthoryear{{Southworth}, {Wheatley} \&
  {Sams}}{{Southworth} et~al.}{2007}]{pjw}
{Southworth} J.,  {Wheatley} P.~J.,    {Sams} G.,  2007, Monthly Notices of the
  Royal Astronomical Society, 379, L11

\bibitem[\protect\citeauthoryear{{Torres}, {Winn} \& {Holman}}{{Torres}
  et~al.}{2008}]{migr}
{Torres} G.,  {Winn} J.~N.,    {Holman} M.~J.,  2008, The Astrophysical
  Journal, 677, 1324

\bibitem[\protect\citeauthoryear{{Vidal-Madjar}, {Lecavelier des Etangs},
  {D{\'e}sert}, {Ballester}, {Ferlet}, {H{\'e}brard} \& {Mayor}}{{Vidal-Madjar}
  et~al.}{2003}]{vidal}
{Vidal-Madjar} A.,  {Lecavelier des Etangs} A.,  {D{\'e}sert} J.-M.,
  {Ballester} G.~E.,  {Ferlet} R.,  {H{\'e}brard} G.,    {Mayor} M.,  2003,
  Nature, 422, 143

\bibitem[\protect\citeauthoryear{{Vidal-Madjar}, {Lecavelier des Etangs},
  {D{\'e}sert}, {Ballester}, {Ferlet}, {H{\'e}brard} \& {Mayor}}{{Vidal-Madjar}
  et~al.}{2008}]{evapagain}
{Vidal-Madjar} A.,  {Lecavelier des Etangs} A.,  {D{\'e}sert} J.-M.,
  {Ballester} G.~E.,  {Ferlet} R.,  {H{\'e}brard} G.,    {Mayor} M.,  2008,
  ArXiv e-prints, 802

\bibitem[\protect\citeauthoryear{{Vilhu}}{{Vilhu}}{1984}]{vil}
{Vilhu} O.,  1984, Astronomy and Astrophysics, 133, 117

\bibitem[\protect\citeauthoryear{{Vilhu} \& {Walter}}{{Vilhu} \&
  {Walter}}{1987}]{vilwil}
{Vilhu} O.,  {Walter} F.~M.,  1987, The Astrophysical Journal, 321, 958

\bibitem[\protect\citeauthoryear{{Wheatley}}{{Wheatley}}{1998}]{xraysatwheatle%
y}
{Wheatley} P.~J.,  1998, Monthly Notices of the Royal Astronomical Society,
  297, 1145

\end{thebibliography}
\bibdata{exoref}
\bibstyle{mn2e}

\label{lastpage}

\end{document}